\documentclass[a4paper,11pt]{article}
\pdfoutput=1 

\usepackage{jheppub} 

\usepackage[T1]{fontenc} 
\usepackage[mathletters]{ucs}
\usepackage[utf8x]{inputenc}
\usepackage{multirow,booktabs,array}
\usepackage{lmodern}
\usepackage{braket}
\usepackage{verbatim}
\usepackage{multirow}
\usepackage{graphics,graphicx}
\usepackage{psfrag}

\usepackage{mathtools}

\newcommand\al{{\alpha}}

\newcommand\Si{\Sigma}

\newcommand\ka{\kappa}
\newcommand\be{\beta}
\newcommand{\Lim}[1]{\raisebox{0.5ex}{\scalebox{0.8}{$\displaystyle \lim_{#1}\;$}}}

\newcommand{\bee}{\begin{equation}}
\newcommand{\ee}{\end{equation}}
\def\ba{\begin{array}}
\def\ea{\end{array}}

\def\bo1{ \left | B^0 (p^+) \right \rangle}

\newcommand{\bea}{\begin{eqnarray}}
\newcommand{\eea}{\end{eqnarray}}

\def\<{ \langle }
\def\>{ \rangle }

\usepackage{amsmath,latexsym,amssymb,slashed}

\title{Correlation functions in the D1-D5 orbifold CFT}

\author{Joan Garcia i Tormo}
\author{and Marika Taylor}
\affiliation{Mathematical Sciences and STAG Research Centre, University of Southampton, \\
Highfield, Southampton, SO17 1BJ, UK.}

\emailAdd{jgt1e15@soton.ac.uk}
\emailAdd{m.m.taylor@soton.ac.uk}

\abstract{The D1-D5 system has an orbifold point in its moduli space, at which it may be described by an ${\cal N} = (4,4)$ supersymmetric sigma model with target space $M^N/S(N)$ where $M$ is $\mathbb{T}^4$ or $K3$. In this paper we consider correlation functions involving chiral operators constructed from twist fields: we find explicit expressions for processes involving a twist $n$ operator joining $n$ twist operators of arbitrary twist. These expressions are universal, in that they are independent of the choice of $M$, and the final results can be expressed in a compact form. We explain how these results are relevant to the black hole microstate programme: one point functions of chiral operators can be used to reconstruct AdS$_3$ near horizon regions of D1-D5 microstates and to match microstates constructed in supergravity with the CFT.}

\begin{document} 
\maketitle
\flushbottom

\section{Introduction}\label{section:introduction}

The D1-D5 system was the original example of black holes in string theory for which microstates could be counted \cite{Strominger:1996sh}. Following the AdS/CFT conjecture \cite{Maldacena:1997re}, the microscopic counting was reinterpreted in terms of the AdS$_3$/CFT$_2$ duality. The D1-D5 black holes under consideration are asymptotically flat but admit a decoupling region that is AdS$_3$, so AdS/CFT is applicable and the black hole microstates are understood as
states of the dual CFT.  

Since the AdS/CFT correspondence is believed to be exact and the dual field theory is unitary, holography implies that the dynamics of such black holes is unitary. This argument shows that there is no information loss but it does not explain exactly how Hawking's arguments \cite{Hawking:1976ra} fail nor does it shed light on the description of black hole microstates at horizon and sub-horizon scales. 

However, AdS/CFT implies much more. The holographic dictionary states that every (stable) state in the conformal field theory should be described by a regular asymptotically AdS geometry; these solutions will generically be stringy in the interior but asymptote to AdS$_3$. Encoded in the AdS asymptotics is information about the dual CFT microstate, expressed in terms of expectation values of operators in that state. Since each of these geometries have the same behaviour near the AdS boundary as the decoupling region of the asymptotically flat black hole, one can re-attach the asymptotically flat region. One thus finds that associated with the black hole there are solutions that look like the black hole up to the horizon scale, but differ from it in the interior: the interior regions are replaced by these asymptotically AdS$_3$ geometries. This is the fuzzball, or black hole microstate, conjecture, originally formulated in the works of Mathur and collaborators 
\cite{Lunin:2001jy,Lunin:2001fv,Lunin:2002qf,Mathur:2002ie,Lunin:2002bj}. Reviews of this conjecture from different perspectives can be found in \cite{Bena:2007kg,Skenderis:2008qn, Mathur:2008wi, Shigemori:2013pqa}.

The D1-D5 system is the most studied setup for the black hole microstate conjecture. While ${\cal N} =4$ SYM is often the prototype AdS/CFT correspondence, the ${\cal N} =4$ SYM microstates corresponding to black holes with macroscopic horizons remain mysterious, with the microstate counting not even being protected by supersymmetric non-renormalisation theorems. However, even though the counting of microstates for D1-D5 black holes is straightforward, there are many challenges in understanding the bulk duals of individual CFT$_2$ microstates. 

There is a long running programme to construct putative black hole microstates as supergravity solutions. This began with the construction of 2-charge D1-D5 microstates \cite{Lunin:2001jy,Lunin:2002bj,Lunin:2002iz,Taylor:2005db,Kanitscheider:2007wq}. In this case the corresponding black hole does not have a macroscopic horizon but the 2-charge case is nonetheless a useful arena to construct microstates explicitly and to make sharp identifications between geometries and the dual CFT states using precision holography \cite{Skenderis:2006ah,Kanitscheider:2006zf,Kanitscheider:2007wq}. 

For three charge microstates, the original examples of supergravity solutions were highly non-generic, with, for example, atypically large angular momentum. Early discussions of (non-generic) microstates for macroscopic black holes can be found in \cite{Mathur:2003hj, Bena:2004tk, Giusto:2004ip,Lunin:2004uu,Jejjala:2005yu,Balasubramanian:2005qu,Berglund:2005vb,Cardoso:2005gj}. Recent constructions of microstates illustrate more generic features, using concepts such as superstrata, see for example \cite{Giusto:2012gt,Giusto:2012jx,Giusto:2012yz,Giusto:2013bda,Bena:2015bea, Chakrabarty:2015foa,Giusto:2015dfa, Bena:2016agb,Bena:2016ypk,Bena:2017geu,Martinec:2017ztd,Bombini:2017got,Bena:2017upb,Bena:2017fvm,Bena:2017xbt,Bossard:2017vii}; there is ongoing study of what fraction of the black hole microstates such constructions can represent.

\bigskip

Several generic lessons from the 2-charge system apply to 3-charge black holes with macroscopic horizons. Firstly, for the supergravity description to be valid, one needs coherent superpositions of microstates in which (single particle) chiral primaries acquire expectation values. The reason is that single particle chiral primaries are dual to supergravity fields; we need the former to acquire expectation values for the interior supergravity geometry of the asymptotically AdS region to carry information about the microstate. 

Secondly, suppose that a given geometry is postulated to be dual to a particular superposition of microstates $| F )$. The expectation values of single particle chiral primaries ${\cal O}_{\Delta}$, of dimension $\Delta$, 
\bee
( F | {\cal O}_{\Delta} | F ) \label{main-vev}
\ee
can then be read off from the asymptotics of the AdS$_3$ region using Kaluza-Klein holography \cite{Skenderis:2006uy}. The matching of not just conserved charges but of whole towers of operators provides very strong evidence for the conjectured duality. This matching was carried out for low dimension operators in the two charge geometries in \cite{Skenderis:2006ah,Kanitscheider:2006zf,Kanitscheider:2007wq} and in three charge geometries in \cite{Giusto:2015dfa}. Holographic four point functions were discussed in \cite{Galliani:2017jlg,Bombini:2017sge}. See also \cite{Skenderis:2006uy,Skenderis:2006di} for an example of matching involving the whole tower of Kaluza-Klein operators - the detailed matching between distributed D3-brane supergravity solutions and the Coulomb branch of ${\cal N} = 4$ SYM. 

The matching of \eqref{main-vev} between the bulk and field theory descriptions relies on being able to compute these expectation values from the dual field theory side and hence, implicitly, uses either non-renormalisation theorems or integrability/bootstrap methods. In the case of the D1-D5 system, one can explicitly compute correlation functions in the orbifold limit of the dual CFT. Correlation functions involving chiral primaries are believed to be non-renormalised as one deforms away from the orbifold point \cite{Pakman:2007hn,Dabholkar:2007ey,Gaberdiel:2007vu,Taylor:2007hs} although the matching between CFT and supergravity is subtle. This non-renormalisation implies that \eqref{main-vev} is non-renormalised away from the orbifold point for 2 charge microstates and it also has implications for expectation values of supergravity operators in 3 charge microstates. 

\bigskip

To develop the precision holography programme for black hole microstates in the D1-D5 system further, one would like to calculate \eqref{main-vev} for general single particle operators from both sides of the correspondence. However, to carry out such calculations from the field theory side, we first need to fill in certain gaps in the literature on correlation functions in the orbifold CFT. 

Single particle chiral primaries of dimension $\Delta$ are described in the orbifold theory in terms of twist operators whose twist is related to the dimension. As we explain in section \ref{section:twist}, to access expectation values for generic dimension operators, we need as a building block amplitudes involving twist $n$ operators joining together $n$ other twist operators. {The goal of this paper is to compute amplitudes for processes in the orbifold CFT involving a twist $n$ operator joining $n$ operators, each of twist $m_i$; we solve this problem for general $n$ and $m_i$ and our results are summarised in \eqref{final0}.  

Correlation functions in the orbifold CFT have been analysed for a variety of applications previously. In what follows here, we will use heavily the pioneering work of Lunin and Mathur on computing three point functions in (super)conformal orbifold theories \cite{Lunin:2000yv,Lunin:2001pw}. An important part of understanding the D1-D5 system is deforming the CFT away from the orbifold limit, and the deformation process has been considered in a number of works \cite{Avery:2009xr,Avery:2010er,Avery:2010vk,Avery:2010hs,Burrington:2012yn,Burrington:2012yq,Burrington:2014yia,Carson:2014ena,Carson:2014yxa,Burrington:2015mfa,Chakrabarty:2015foa,Carson:2015ohj,Carson:2016uwf,Burrington:2017jhh,Carson:2017byr,Burrington:2018upk}. The methodology used here is closely related to that of \cite{Carson:2014ena,Carson:2014yxa}, although our motivations are somewhat different. Efficient methods for computing certain extremal correlation functions in the orbifold theory and  relations with spin chains were presented in \cite{Pakman:2009zz,Pakman:2009ab,Pakman:2009mi}. 

\bigskip

The plan of this paper is as follows. In section \ref{section:translation} we review key features of the D1-D5 orbifold SCFT. In section \ref{section:twist} we explain why processes involving a twist $n$ operator joining $n$ other operators are essential to calculating expectation values \eqref{main-vev} from the field theory side. Section \ref{section:vev} contains the technical computation of the amplitude for such processes. In section \ref{section:conc} we conclude by discussing implications and applications of our results. 

\pagebreak

\section{Review of key features of the D1-D5 orbifold CFT }\label{section:translation}

In this section we review essential features of the D1-D5 orbifold CFT; a more complete review of the D1-D5 system 
can be found in \cite{David:2002wn}. Consider type IIB string theory compactified on $X \times S^1$, with $X$ being $\mathbb T^4$ or $K3$. 
Let $N_5$ D5-branes wrap the five compact dimensions and $N_1$ D1-branes wrap the $S^1$.
We take $X$ to be string scale and assume that the scale of the $S^1$ is much larger.
D1-D5 black hole solutions in the supergravity limit are asymptotic to $M^{4,1}\times S^1 \times X$.
The geometry of the decoupled near horizon limit is AdS$_3\times S^3 \times X$, and there is supersymmetry enhancement (see \cite{Boonstra:1997dy} and references therein).

The CFT dual to the decoupling region geometry is a two-dimensional superconformal CFT. In what follows, we will focus on the theory for $X = \mathbb T^4$,
although much of our later analysis also holds for $K3$, i.e. it does not rely on features specific to $\mathbb T^4$. For toroidal compactifications, the SCFT is an
$\mathcal N = (4,4)$ superconformal sigma model with central charges $c = \bar c = 6 N_1 N_5$; this theory can be viewed as a deformation of a free orbifold CFT with target space $(\mathbb T^4)^{N_1 N_5}/S( N_1 N_5)$, where $S(n)$ is the symmetric group.

In this paper we will be interested in calculating correlation functions involving chiral primary operators in this theory. Note that the three point functions of chiral primaries themselves are protected (see \cite{Pakman:2007hn,Dabholkar:2007ey,Gaberdiel:2007vu,Taylor:2007hs}) and will therefore agree with the corresponding three point functions calculated in supergravity. 
The correlation functions we calculate can also be used as building blocks for correlation functions involving less supersymmetric operators. 
We will work in Euclidean signature on a cylinder which is parameterised as
\bee
w = \tau + i \sigma 
\ee
where $0 \le \sigma < 2 \pi$ and $ -\infty < \tau < \infty$. 

\bigskip
Let us now briefly review the orbifold field theory description.
The Hilbert space of the orbifold theory decomposes into twisted sectors which are labelled by conjugacy classes of the symmetry group $S( N_1 N_5)$ which consists of cyclic groups of various lengths; see discussion in \cite{Jevicki:1998bm}. The conjugacy classes that occur and their multiplicity are subject to the constraint
\bee
\sum_i n_i m_i = N
\ee
where $m_i$ is the length of the cycle (the twist), $n_i$ is the multiplicity of the cycle and $N = N_1 N_5$. There is a direct correspondence between the conjugacy classes and the long/short string picture of the D1-D5 system \cite{Maldacena:1996ds}. 
The symmetry group of the SCFT is $SU(1,1|2) \times SU(1,1|2)$, as is discussed in the review \cite{David:2002wn}.  
The symmetry that is identified with the $SO(4)_E$ isometry of the $S^3$ in the geometry is the $SO(4)$ R-symmetry in the $\mathcal{N}=(4,4)$ superconformal algebra. The SCFT has another $SO(4)$ symmetry which is identified with the $SO(4)_I$ of the torus, when $X = \mathbb T^4$.

Chiral primaries can be precisely described at the orbifold point; they are associated with the cohomology of $X$. 
The NS sector chiral primaries can be labeled as ${\cal O}_m^{(p,q)}$ where $m$ is the twist and $(p,q)$ labels the associated cohomology class. The weights $(h, \tilde{h})$ and R charges $(j_3, \tilde{j}_3)$ of such chiral primaries are given by
\bee
h^{\rm NS} = j_3^{\rm NS} = \frac{1}{2} (p + m -1); \qquad
\tilde{h}^{\rm NS} = \tilde{j}_3^{\rm NS} = \frac{1}{2} (q + m -1). \label{NS-charge}
\ee
The complete set of chiral primaries is then built from products of the form
\bee
\prod_l ({\cal O}^{(p_l,q_l)}_{m_l})^{n_l} \qquad 
\sum_l n_l m_l = N
\ee
with symmetrisation over $N$ copies of the CFT implicit. 

Spectral flow maps chiral primaries in the NS sector to R ground states for which
\bee
h^{\rm R} = h^{\rm NS} - j_3^{\rm NS} + \frac{c}{24} \qquad j_3^{\rm R} = j_3^{\rm NS} - \frac{c}{12}
\ee
where $c$ is the central charge of the CFT and analogous expressions hold for the right moving sector. Each of the NS sector chiral primaries is thus mapped by spectral flow to a Ramond ground state operator 
\bee
\prod_l ({\cal O}^{(p_l,q_l)}_{m_l})^{n_l} \rightarrow \prod_l ({\cal O}^{{\rm R}(p_l,q_l)}_{m_l})^{n_l}
\ee
of definite R charge
\bee
j_3^{\rm R} = \frac{1}{2} \sum_l (p_l - 1) n_l \qquad
\tilde{j}_3^{\rm R} = \frac{1}{2} \sum_l (q_l - 1) n_l.
\ee
Note that the Ramond operators obtained from primaries associated with the $(1,1)$ cohomology have zero R charge. 

\bigskip

The Ramond ground states are the microstates associated with the 2-charge D1-D5 black hole; the associated entropy is
\bee
S = 2 \pi \sqrt{\frac{C(X) N}{6}}
\ee
where $C(X)$ is determined by the cohomology (and $C = 12$ and $24$ for K3 and $\mathbb{T}^4$). The corresponding black holes do not have macroscopic horizons. The famous 3-charge black holes with macroscopic horizons discussed in \cite{Strominger:1996sh} correspond to exciting the left moving sector with momentum $P$; the resulting entropy is then
\bee
S =  2 \pi \sqrt{\frac{C(X) N P}{6}}
\ee
(where implicitly we assume that $P \gg N$). We will not need the detailed description of these microstates in this paper but all such states can be viewed as excitations over the  ground states i.e. the generic structure is 
\bee
{\cal O}_{P} \prod_l ({\cal O}^{{\rm R}(p_l,q_l)}_{m_l})^{n_l}
\ee
where ${\cal O}_P$ describes the excitation of momentum $P$. As discussed in early works such as \cite{Maldacena:1996ds}, most of the 3-charge microstates are associated with excitations over maximal and near maximal twist ground states (``long strings'') as there are more ways to fractionate the momentum over such states.

\subsection{Explicit description in terms of free fields}

In what follows we will for the most part not need to use an explicit description in terms of free fields. In this section we therefore 
include only the aspects of the free field description that are needed for the correlation function calculation; further details can be found in   
\cite{David:2002wn}. Our notation follows closely that of \cite{Giusto:2015dfa} and \cite{Bena:2016agb}.


Let us first introduce appropriate notation.  We denote the $SO(4)$ symmetry associated with the torus as $SU(2)_{\mathcal C} \times SU(2)_{\mathcal A}$, and use the labels $\al,\dot{\al}$ for the left and right R-symmetry, $\dot{A}$ for $SU(2)_{\mathcal C}$ and $A$ for $SU(2)_{\mathcal A}$.
We will use $\{+, -\}$ for the $\al$ index, $\{\dot +, \dot -\}$ for the $\dot \al$ index, $\{1, 2\}$ for $A$ and $\{\dot 1, \dot 2\}$ for $\dot A$.
A subindex $(r)$, which runs from 1 to $N_1 N_5$, is used to label the copies of the torus.
Fields and operators corresponding to the right  moving sector are denoted with a tilde.

At the orbifold point, the CFT has free fields
\begin{equation}
\left(X^{\dot A A}_{(r)}(w, \bar{w}), \psi^{\al\dot A}_{(r)}(w), \tilde{\psi}^{\dot{\al} \dot A}_{(r)}(\bar{w})\right),
\end{equation}
that is, four bosons and four doublets of fermions.
The mode expansion of the fermions in the Ramond sector is then
\begin{equation}
\psi^{\al\dot A}_{(r)}(w) = \sum_{n\in\mathbb Z} \psi^{\al\dot A}_{n(r)} e^{- n w}, \qquad \tilde{\psi}^{\dot{\al} \dot A}_{(r)}(\bar{w}) = \sum_{n\in\mathbb Z} \tilde{\psi}^{\dot{\al} \dot A}_{n(r)}e^{- n \bar{w}}.
\end{equation}
The Hermitian properties of the fermions are $\psi^{+\dot 1 \dagger}_{n(r)} = -\psi^{-\dot 2}_{-n(r)}$, $\psi^{+\dot 2 \dagger}_{n(r)} = \psi^{-\dot 1}_{-n(r)}$, and similarly for the right-moving sector.
The Ramond vacuum state $\ket{++}_{(r)}$ is defined by
\begin{equation}
\psi^{+\dot 1}_{0(r)}\ket{++}_{(r)} = \psi^{+\dot 2}_{0(r)}\ket{++}_{(r)} = 0, \qquad \tilde{\psi}^{\dot +\dot 1}_{0(r)}\ket{++}_{(r)} = \tilde{\psi}^{\dot +\dot 2}_{0(r)}\ket{++}_{(r)} = 0,
\label{++def}
\end{equation}
and the R-symmetry currents can be expressed explicitly as 
\begin{equation}
J^{\al\be}_{(r)}(w) = \frac 1 2 \psi^{\al\dot A}_{(r)}(w)\epsilon_{\dot A \dot B}\psi^{\be\dot B}_{(r)}(w), \qquad \tilde{J}^{\dot{\al}\dot{\be}}_{(r)}(\bar{w}) = \frac 1 2 \tilde{\psi}^{\dot{\al}\dot A}_{(r)}(\bar{w})\epsilon_{\dot A \dot B}\tilde{\psi}^{\tilde{\be}\dot B}_{(r)}(\bar{w}),
\end{equation}
where the operators are normal-ordered with respect to the $\ket{++}_{(r)}$ ground state.
The generators are $J^3_{(r)} := -J^{+-}_{(r)} + \frac 1 2$, $J^+_{(r)} := J^{++}_{(r)}$ and $J^-_{(r)} := -J^{--}_{(r)}$.
We define the other ground states with their zero modes,
\begin{equation}
\ket{-+}_{(r)} = J^-_{0(r)}\ket{++}_{(r)}, \qquad \ket{+-}_{(r)} = \tilde{J}^-_{0(r)}\ket{++}_{(r)},
\label{--def}
\end{equation}
and the (0,0) spin one (under ($J^3_{(r)}, \tilde{J}^3_{(r)}$)) as
\begin{equation}
\ket{00}_{(r)} := \lim_{z\to 0}\mathcal O^{-\dot -}_{00(r)} \ket{++}_{(r)} = \frac{1}{\sqrt 2}\psi^{-\dot A}_{0(r)}\epsilon_{\dot A\dot B}\tilde{\psi}^{\dot -\dot B}_{0(r)}\ket{++}_{(r)},
\label{00def}
\end{equation}
where we have used the conformal map $z = e^{w}$.

The operator that induces a cyclic permutation of $\ka\geq 2$ copies of elementary fields is denoted by $\Si^{\al \dot \al}_\ka$.
It generates the twisted states, the cycles of length $\ka$; in other words, it joins $\ka$ strings of winding one into a single string of winding $\ka$.
We define strands of length $\ka$ as
\begin{equation}
\ket{++}_\ka := \lim_{z\to 0}|z|^{\ka - 1}\Si^{-\frac{\ka - 1}{2}, -\frac{\ka - 1}{2}}_\ka (z, \bar z)\prod_{r = 1}^\ka \ket{++}_{(r)},
\label{strandk}
\end{equation}
where $\Si^{-\frac{\ka - 1}{2}, -\frac{\ka - 1}{2}}_\ka$ is the lowest weight state in the $\Si_\ka$ multiplet.
It has spin (-$\frac{\ka - 1}{2}$, -$\frac{\ka - 1}{2}$), and these are the superindices that we have written, to make the spin explicit -- they are not the $\al \dot \al$ indices of the operator, and that is why the second does not have a dot.
Acting on the $\ka$ copies of the ground state generates the state $\ket{++}_\ka$ in the Ramond sector, which has spin ($\frac 1 2$, $\frac 1 2$) and winding $\ka$.

To obtain independent fields we need to diagonalise the boundary conditions using combinations
\begin{equation}
\psi^{\al\dot A}_{\rho}(z) = \frac{1}{\sqrt{\ka}}\sum_{r = 1}^{\ka} e^{-2\pi i\frac{r\rho}{\ka}} \psi^{\al\dot A}_{\rho (r)}(z), \qquad \mathrm{with} \qquad \rho = 0, 1, ..., \ka - 1.
\label{diagonalfermions}
\end{equation}
The total R charge of a strand (which is $\sum^\ka_{r = 1}J^3_{(r)}$) can then be changed using the zero modes of $\psi^{\al\dot A}_{ 0}$.
Hence,
\begin{equation}
\ket{-+}_\ka = J^-_{0 \rho = 0}\ket{++}_\ka, \qquad \ket{+-}_\ka = \tilde{J}^-_{0 \rho = 0}\ket{++}_\ka.
\end{equation}
For the spin zero ground states we act with the zero mode of $\sum_r \mathcal O^{2\dot 2}_{(r)}$,
\begin{equation}
\ket{00}_\ka = -\frac{i}{\sqrt 2} \psi^{- \dot A}_{0 \rho=0} \epsilon_{\dot A\dot B}\tilde{\psi}^{\dot -\dot B}_{0 \rho=0} \ket{++}_\ka.
\end{equation}
This gives explicitly constructions for all ground states associated with even cohomology classes, which is all we will consider here. The correspondence between these ground states and cohomologies is
\bee
{\cal O}^{R (2,2)}_{\kappa} \leftrightarrow \ket{ + +}_\ka \qquad
{\cal O}^{R (1,1)}_{\kappa} \leftrightarrow \ket{ 0 0 }_\ka \qquad
{\cal O}^{R (0,0)}_{\kappa} \leftrightarrow \ket{ - - }_\ka 
\ee
and similarly
\bee
{\cal O}^{R (2,0)}_{\kappa} \leftrightarrow \ket{ + -}_\ka \qquad
{\cal O}^{R (0,2)}_{\kappa} \leftrightarrow \ket{ - +}_\ka
\ee
In the following sections we will revert to the notation using cohomology, as this notation reflects the geometry of the target space and moreover illustrates more naturally the connections with the corresponding supergravity operators.

Given these Ramond ground states, one can add momentum excitations by, for example, acting with raising modes of the currents 
\bee
(J^+_{-\frac{n_{\kappa}}{\kappa}})^{m_{\kappa}} \ket{00}_{\ka} \label{3cm}
\ee
with $n_{\kappa}$ and $m_{\kappa}$ integers. The excitations should be 
such that the total momentum added to the state is integral. Clearly the larger the twist, the greater the number of possibilities for distributing the excitations, and this is why most of the 3-charge microstates are associated with long strings. The specific states corresponding to superstrata solutions are discussed in 
\cite{Bena:2015bea, Chakrabarty:2015foa,Giusto:2015dfa, Bena:2016agb,Bena:2016ypk}. 


\section{Twist operator amplitudes} \label{section:twist}

In this paper we will focus on processes in which a twist operator of twist $n$ joins $n$ operators of twists $(m_1,m_2,\cdots,m_n)$, to form a single operator of twist $M := (m_1 + \cdots m_n)$. Following the usual correspondence between twist operators and strings, we illustrate these operators via strings with winding equivalent to the twist. The process of interest is illustrated in figure~\ref{fig:1}.

\begin{figure}[tbp]
\centering
\includegraphics[width=0.5\linewidth]{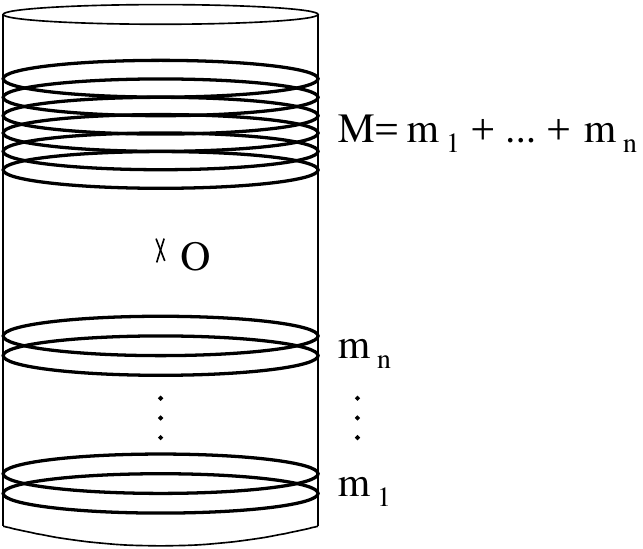}
\caption{Joining of $n$ component strings to form one single string.}
\label{fig:1}
\end{figure}

Let us consider the following example of such a process, expressed first in NS sector language. Let each component string be generated by a chiral primary i.e.
\bee
{\cal O}^{(p_i,q_i)}_{m_i} | 0 \rangle_{m_i}.
\ee
Let the operator of twist $n$ joining these component strings be a primary, which we denote as 
\bee
{\cal O}^{ (p,q)}_{n}.
\ee
The joining process then acts as
\bee
{\cal O}^{ (p,q)}_{n} | \prod_{i=1}^n {\cal O}^{(p_i,q_i)}_{m_i}  \rangle \rightarrow | \chi \rangle.
\ee
Here the state $| \chi \rangle$ by construction has R charges 
\bee
j_3^{NS} = \frac{1}{2} \left (p + \sum_{i=1}^n p_i  \right ) + \frac{1}{2} M - \frac{1}{2} \qquad
\tilde{j}_3^{NS} =  \frac{1}{2} \left (q + \sum_{i=1}^n q_i \right  ) + \frac{1}{2} M - \frac{1}{2}. 
\ee
We begin by considering processes in which 
\bee
\left (p + \sum_{i=1}^n p_i  \right ) = P \le 2 \qquad
\left (q + \sum_{i=1}^n q_i  \right ) = Q \le 2
\ee
for which 
\bee
| \chi \rangle = e^{ {\cal O}_{c}} | {\cal O}^{(P,Q)}_{M}  \rangle
\ee
where $e^{ {\cal O}_{c}}$ describes coherent excitations of zero R charge over the NS chiral primary. 

By construction the one point function
\bee
\langle {\cal O}^{(P,Q)}_{M}| {\cal O}^{ (p,q)}_{n} (I)^{M -n} | \prod_{i=1}^n {\cal O}^{(p_i,q_i)}_{m_i}  \rangle \label{3cf}
\ee
is thus (generically) non-zero. In this expression we include explicitly the factor of $(M-n)$ copies of the identity operator, to emphasise
that this correlation function is computed in $M$ copies of the CFT. 

\bigskip

The one point function \eqref{3cf} is not a ``physical'' one point function in the orbifold CFT as we have imposed neither $M=N$ nor symmetrisation over copies of the CFT, and we have not specified the full state. However, \eqref{3cf} is an important building block for physical one point functions of interest in the context of D1-D5 holography. For example, the one point function contribution \eqref{3cf} is relevant to the computation of one point functions of single trace chiral primary operators in Ramond ground states. Spectral flow of \eqref{3cf}
gives
\bee
\langle {\cal O}^{R (P,Q)}_{M}| {\cal O}^{R (p,q)}_{n}  | \prod_{i=1}^n {\cal O}^{R (p_i,q_i)}_{m_i} \rangle.\label{1cf1}
\ee
The basis of Ramond ground states in the orbifold CFT was described previously. Since each Ramond ground state is an eigenstate of $J^3$ and $\tilde{J}^3$, the expectation of a single trace chiral
primary (with non zero R charge) is necessarily zero. However, R ground states that admit holographic supergravity duals can be expressed in terms of projections of coherent 
superpositions \cite{Kanitscheider:2006zf,Kanitscheider:2007wq} i.e. as
\bee
| {\cal O}^R_{c} \rangle = \sum_A c_A  | {\cal O}^{R}_A \rangle \label{coher}
\ee 
where $A$ labels the complete set of Ramond ground states and the coefficients $c_A$ are inherited from projections of coherent superpositions. 

More precisely, there is a direct correspondence between the curves describing the holographic supergravity solutions and these coherent superpositions, described in detail in \cite{Kanitscheider:2006zf,Kanitscheider:2007wq}. In brief, the curves ${\cal F}(v)$ describing the supergravity solutions can be decomposed into Fourier modes
\bee
{\cal F}(v) = \sum_{n > 0} \frac{1}{\sqrt{n}} \left ( \alpha_{n} e^{- in v} + \alpha_{n}^{\ast} e^{inv} \right ).
\ee
Now introduce auxiliary harmonic oscillators as operators, $\hat{a}_{n}$, and define coherent states associated with these operators as 
\bee
\hat{a}_n \left . | \alpha_n \right )  = \alpha_n \left . | \alpha_n \right ).
\ee 
There is thus a coherent state associated with the curve 
\bee
\left . | {\cal F} \right ) = \prod_n \left . | \alpha_n \right )
\ee
The coherent states can be expressed in terms of Fock states in the standard way as 
\bee
\left .  | \alpha_n \right ) = \exp \left ( - \frac{ | \alpha_n |^2}{2}  \right ) \sum_k \frac{\alpha_n^k}{k!} (\hat{a}_n^{\dagger})^k \left .  | 0 \right \rangle
\ee
and we can then project from $\left |  {\cal F} \right )$ the Fock states that satisfy the constraint
\bee
\prod (\hat{a}_{n_l}^{\dagger})^ {m_l} | 0 \rangle \qquad \sum_l n_l m_l = N_1 N_5. 
\ee
The final step is to retain only these terms from $\left . | {\cal F} \right )$ and map the auxiliary harmonic oscillators to CFT R operators. The Ramond ground state operators are in one-to-one correspondence with the cohomology of the target space for the orbifold CFT; thus the number of independent curves defining the supergravity  geometries is given by the sum of the Hodge numbers of this target space. The result indeed gives a linear superposition of Ramond ground states \eqref{coher} with superposition coefficients $c_A$ inherited from the defining curves. 

An important feature of the superposition \eqref{coher} is that it is not in general an eigenstate of R symmetry. This implies that charged operators can acquire expectation values in this state. One can extract these expectation values from the supergravity solutions via holographic renormalisation. For the D1-D5 ground states, non-renormalisation theorems are believed to exist, implying that these expectation values match between supergravity and the CFT in the orbifold limit (although the required matching between supergravity and CFT operators is subtle \cite{Taylor:2007hs}). 

Hence in a generic superposition the expectation of a single trace operator is 
\bee
\langle {\cal O}_c | {\cal O}^{p,q}_n   | {\cal O}_{c} \rangle = \sum_{A,B} c_A^{\ast} c_B \langle {\cal O}^{R}_A | {\cal O}^{p,q}_n | {\cal O}^R_B\rangle. \label{bps1}
\ee
It is now apparent that \eqref{1cf1} is a building block for computing such one point functions: non-vanishing terms in this one point function are associated with the twist $n$ operator joining component strings. 

Computation of \eqref{bps1} for general twist $n$ operators would allow precision holography for two charge microstates to be tested further, using the methods of  \cite{Skenderis:2006ah,Kanitscheider:2006zf,Kanitscheider:2007wq}. A good understanding of \eqref{bps1} is also needed to calculate one point functions of supergravity operators (single particle chiral primaries) in three charge microstates. A typical 3-charge microstate is built out of superpositions of Ramond ground states excited by left moving momenta as in \eqref{3cm}. The one point functions will then reduce to sums of amplitudes of the type
\bee
\langle {\cal O}_{P_A} {\cal O}^{R}_A | {\cal O}^{p,q}_n | {\cal O}_{P_B}{\cal O}^R_B \rangle
\ee
where ${\cal O}_{P_A}$ and ${\cal O}_{P_B}$ denote the operators exciting left moving momenta over the ground states.  For excitations such as \eqref{3cm} one can then use commutation relations to reduce this calculation to \eqref{bps1}; this will be discussed in future work. Note that it is the one point functions of single particle chiral primaries (single strings) that are of most interest in matching holographically with microstate geometries, as it is these values that are captured by the asymptotics of the interior AdS$_3$ regions.

\section{Computation of twist operator expectation value} \label{section:vev}

In this section we will focus on the computation of 
\bee
\langle {\cal O}^{R (P,Q)}_{M}| {\cal O}^{R (p,q)}_{n}  | \prod_{i=1}^n {\cal O}^{R (p_i,q_i)}_{m_i} \rangle.\label{1cf}
\ee
Our methods will follow the approach pioneered by \cite{Lunin:2000yv,Lunin:2001pw} and used in the case of a twist two operator in 
\cite{Carson:2014yxa,Carson:2014ena}.

The one point function is computed on the cylinder, with the operator inserted at the location $w_0$ i.e. the explicit computation that is required is 
\bee
\langle {\cal O}^{R (P,Q)}_{M}| {\cal O}^{R (p,q)}_{n}  (w_0) | \prod_{i=1}^n {\cal O}^{R (p_i,q_i)}_{m_i} \rangle.\label{1cf2}
\ee
Note that this one point function should be independent of the insertion point $w_0$: the Ramond ground states are eigenstates of $L_0$ and $\bar{L}_0$ and thus one can freely use translations to move the insertion point around the cylinder. 
In practice we calculate this one point function by lifting to a covering space, and computing the appropriate $(n+2)$ point function. We begin by discussing the required maps to covering spaces. 

\subsection{Maps to covering space}

We begin by working on a cylinder with coordinate $w$. The cylinder is mapped to the complex plane using the standard exponential map $z = \exp(w)$. The CFT fields are however multi-valued on the $z$ plane due to the presence of the twist fields. We thus map to a covering space with coordinate $t$ where the fields are single valued. The twist operators are punctures on the $t$ plane. 

We can regulate the single component string insertions using the following map from the $z$ plane to the $t$ plane, in analogy to the map used in \cite{Carson:2014yxa,Carson:2014ena}:
\bee
z = t^{m_1} (t - a_2)^{m_2}(t - a_3)^{m_3} \cdot ... \cdot (t - a_n)^{m_n}. \label{map1}
\ee
On the cylinder the initial components strings are at $w \rightarrow - \infty$, which corresponds to $z=0$ on the $z$ plane. On the $t$ plane a string of winding $m_i$ is mapped to position $a_i$; we set $a_1  = 0$ for simplicity, without loss of generality in what follows. 
The final component string is at $w \rightarrow \infty$ on the cylinder, which maps to $t \rightarrow \infty$ on the plane. The twist $n$ operator 
is inserted at $w_0$ on the cylinder which corresponds to $\exp(w_0)$ on the $z$ plane. 

A priori the parameters $a_i$ are not fixed in terms of the original parameter $w_0$ on the cylinder. However, the ramification map should be such that $dz/dt$ has a zero of order $(n-1)$ at the location of the twist $n$ operator. Let $t_0$ be the location of the twist $n$ operator; then 
\bee
\frac{dz}{dt} = (t - t_0)^{n-1} P_{M-n} (t) \label{double}
\ee
with $P_{M-n}(t)$ a polynomial of order $(M-n)$ with no zero at $t_0$. 
We can understand this as follows. The map \eqref{map1} is a polynomial of order $M$ with $M$ non-distinct zeros: it has a zero of order $m_a$ at $t=0$ and so on. Thus its first derivative is a polynomial of order $(M-1)$.  Now $dz/dt$ has a total of $(M-n)$ zeros at locations $a_i$: it has a zero of order $(m_1 - 1)$ at $t=0$, a zero of order $(m_2 - 1)$ at $t=a_2$ etc. By the fundamental theory of algebra, $dz/dt$ has an additional $(n-1)$ zeros, and these are located at the position of the twist $n$ operator. 

The original map \eqref{map1} has $(n-1)$ parameters $(a_2, ..., a_n)$. These parameters are determined by the condition that $dz/dt$ has a zero of order $(n-1)$ at the location $t_0$. Note that $t_0$ is related to the original insertion point on the cylinder via the map
\bee
\exp(w_0) = t_0^{m_1} (t_0 - a_2(t_0))^{m_2}(t_0 - a_3(t_0))^{m_3} \cdot ... \cdot (t_0 - a_n(t_0))^{m_n} \label{wt-rel}
\ee
where here we indicate that the positions $a_i$ can be expressed as functions of $t_0$. 

\bigskip

Let us first illustrate these general discussions in the context of $n=3$; the case of $n=2$ is discussed in detail in \cite{Carson:2014yxa,Carson:2014ena}. For $n=3$ the ramification map is
\bee
z = t^{m_1} (t  - a_2)^{m_2} (t - a_3)^{m_3}
\ee
and thus 
\bee
\frac{dz}{dt} = t^{m_1 -1} (t - a_2)^{m_2  - 1} (t - a_3)^{m_3 - 1} \left ( m_1 (t - a_2) (t-a_3) + m_2 t (t-a_3) + m_3 t (t - a_2) \right ).
\ee
The requirement that this takes the form \eqref{double} imposes
\bee
a_2 = \bar{a}_2 t_0 \qquad
a_3 = \bar{a}_3 t_0
\ee
where $(\bar{a}_2, \bar{a}_3)$ satisfy
\bee
\bar{a}_2 \bar{a}_3 = \frac{M}{m_1}; \qquad 
\bar{a}_2 \left ( 1 - \frac{m_2}{M} \right ) + \bar{a}_3 \left ( 1 - \frac{m_3}{M} \right ) = 2. 
\ee
These equations can be solved to give
\bee
\bar{a}_2  \left ( 1 - \frac{m_2}{M} \right )  = 1 \pm i \sqrt{\frac{m_2 m_3}{m_1 M}} \qquad
\bar{a}_3  \left ( 1 - \frac{m_3}{M} \right )  = 1 \mp i \sqrt{\frac{m_2 m_3}{m_1 M}}.
\ee
With these solutions we can relate $w_0$ and $t_0$ as
\bee
\exp(w_0) = t_0^M (1 - \bar{a}_2)^{m_2} (1 - \bar{a}_3)^{m_3}. 
\ee
Clearly the relation between $t_0$ and $w_0$ is not unique; we will clarify this issue below in the case of general $n$. 

\bigskip 

We can now immediately generalise to arbitrary $n \ge 2$. The ramification map is 
\bee
z = t^{m_1} \prod_{i=2}^{n} (t - a_i)^{m_i}
\ee
and the requirement that $t=t_0$ is a zero of $dz/dt$ of order $(n-1)$ \eqref{double} imposes $(n-1)$ relations on the $a_i$: $a_i = \bar{a}_i t_0$ with
\bee
\prod_{i=2}^n \bar{a}_i = \frac{M}{m_1}   \qquad n \ge 2 
\ee
together with $(n-2)$ further conditions 
\bea
\sum_{i=2}^{n} \bar{a}_i \left (1 - \frac{m_i}{M} \right ) &=& (n -1) \qquad n  \ge  3  \nonumber \\
 \sum_{i=1}^n m_i  \sum_{l \neq n \neq i} \bar{a}_n \bar{a}_l &=& M \frac{(n-1) (n-2)}{2}  \qquad n \ge  4 
\eea
and so on. For example, for $n \ge 5$ we would in addition need the cubic relation between the $\bar{a}_i$. Note that for $n=2$ we can immediately read off $\bar{a}_2 = M/m_1$ from the expression above, which is in agreement with the ramification map used in  \cite{Carson:2014yxa,Carson:2014ena}.

In analogy to the $n = 3$ case, it is natural to write the solutions of these equations as 
\bee
\left (1 - \frac{m_i}{M} \right ) \bar{a}_i = \left (1  + \bar{a} \exp(i \phi_i) \right )  \label{sol2}
\ee
where the phases $\phi_i$ satisfy
\bee
\prod_{i=2}^n \exp( i \phi_i) = 1 \qquad n \ge 3 \label{phase1}
\ee
and
\bea
\sum_{i=2}^n \exp (i \phi_i) &=& 0 \qquad n \ge 3 \label{phase2}  \nonumber \\
\sum_{i \neq j} \exp (i (\phi_i + \phi_j) ) &=& 0 \qquad n \ge 4 \nonumber \\
\sum_{i \neq j \neq k} \exp(i (\phi_i + \phi_j + \phi_k) ) &=& 0 \qquad n \ge 5,
\eea
and so on. Solutions for these phases are:
\bea
(n-1) \in 2 Z : \quad \phi_i &=& \frac{(i-1) \pi}{n-1}, \quad  \phi_{i+1}  = - \frac{ (i-1) \pi}{n-1}, \quad i \in 2 Z, \quad i \ge 2 \label{phase-sol} \\
n \in 2 Z : \quad \phi_2 &=& 0, \quad  \phi_i = \frac{(i-1) \pi}{n-1}, \quad  \phi_{i+1}  = - \frac{(i-1) \pi}{n-1}, \quad (i - 1) \in 2 Z, \quad  i \ge 3. \nonumber
\eea
Note that these solutions are not unique i.e. any permutation of the phases will also solve the equations. One can also shift all of the phases by an equal amount i.e. $\phi_i \rightarrow \tilde{\phi}_i = \phi_i + \lambda$, satisfying \eqref{phase2} but now instead of \eqref{phase1} one has
\bee
\prod_{i=2}^n \exp( i \phi_i) = \exp( i (n-1) \lambda). 
\ee
This shift can trivially be absorbed into the parameter $\bar{a}$  in \eqref{sol2} and thus we can always set $\lambda = 0$ without loss of generality. 

The parameter $\bar{a}$ in \eqref{sol2} satisfies
\bee
1 + \bar{a}^{n-1} = \frac{M}{m_1}  \prod_{i=2}^n \left ( 1 - \frac{m_i}{M} \right ).
\ee
We can solve this equation as follows. First note that
\bee
m_1 = M - \sum_{i=2}^n m_i
\ee
and introduce the notation $\nu_i := m_i/M$, where clearly $0 < \nu_i < 1$. Then 
\bee
\bar{a}^{n-1} = \frac{ \prod_{i=2}^n \left ( 1 - \nu_i \right )  }{\left (1 - \sum_{i=2}^n \nu_i \right )}  - 1.
\ee
Now for $n > 2$
\bee
\prod_{i=2}^n \left ( 1 - \nu_i \right ) > \left (1 - \sum_{i=2}^n \nu_i \right ). \label{id3}
\ee
This follows from induction: if one assumes that the identity holds for $n$ then for $(n+1)$
\bea
\prod_{i=2}^{n+1} \left ( 1 - \nu_i \right ) &=& \left (1 - \nu_{n+1} \right ) \prod_{i=2}^n \left ( 1 - \nu_i \right ) \nonumber \\
& > & 
\left (1 - \nu_{n+1} \right ) \left (1 - \sum_{i=2}^n \nu_i \right ) >  \left (1 - \sum_{i=2}^{n+1} \nu_i \right ).
\eea
The identity is true for $n=3$ as 
\bee
\left ( 1 - \nu_2 \right ) \left ( 1 - \nu_3 \right ) > \left ( 1 - \nu_2 \nu_3 \right)
\ee
and therefore by induction \eqref{id3} holds for all $n \ge 3$. 

Hence we may write 
\bee
\bar{a}^{n-1} =:  Q = \frac{1}{\nu_1}  \prod_{i=2}^n \left ( 1 - \nu_i \right ) - 1 \label{bar-a}
\ee
where $Q \in Q^{+}$. The $(n-1)$ roots of this equation are 
\bee
\bar{a} = Q^{\frac{1}{n-1}} \exp \left ( \frac{2 \pi i k }{n-1} \right  ), \label{bar-a2}
\ee
with $k = 0, 1, ..., (n-2)$. We can however fix $k=0$ so that $\bar{a}$ is real: 
other choices of $k$ are equivalent to rotations of the phases $\phi_i$. 

\bigskip

Thus for general $n$ we have concluded that the map between $w_0$ and $t_0$ \eqref{wt-rel} takes the form
\bee
\exp(w_0) = t_0^{M} \prod_{i=2}^n (1 - \bar{a}_i)^{m_i} \label{wtmap}
\ee
where
\bee
\bar{a}_i = \frac{ \left ( 1 + \bar{a} \exp (i \phi_i) \right ) }{\left ( 1 - \nu_i \right )} \label{bari}
\ee
with $\bar{a}$ given by \eqref{bar-a2} and the phases $\phi_i$ are given by \eqref{phase-sol}. 
It is useful to illustrate the structure of this ramification map as follows. If we consider the combinations 
\bee
A_i = \bar{a}_i (1 - \nu_i) 
\ee
then the $A_i$ are located at the vertices of a regular $n$-sided polygon, with centre one, as shown in Figure~\ref{fig:2}. 

\begin{figure}[tbp]
\centering
\includegraphics[width=0.5\linewidth]{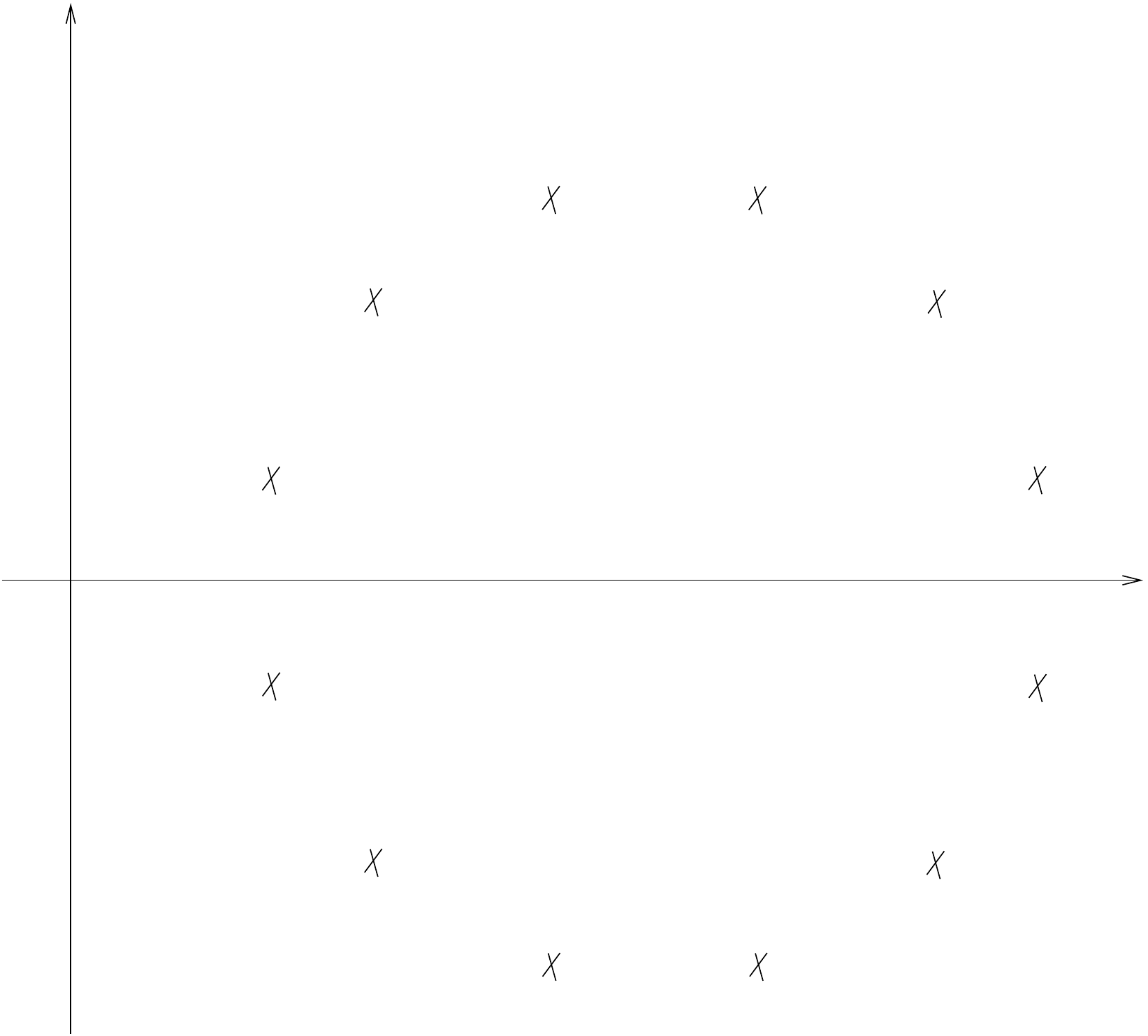}
\caption{Illustration of structure of ramification map; the crosses denote $A_i$.}
\label{fig:2}
\end{figure}

\subsection{Computation of one point function}

In this section we explain how the required one point function \eqref{1cf2} can be computed in terms of a correlation function in the $t$ plane. The methodology follows closely \cite{Carson:2014yxa,Carson:2014ena}, which in turn exploited the techniques for computing orbifold CFT correlation functions developed in 
\cite{Lunin:2000yv,Lunin:2001pw}.

The one point function \eqref{1cf2} is calculated by first lifting to the $z$ plane to give:
\bee
\langle {\cal O}^{R (P,Q)}_{M}| {\cal O}^{R (p,q)}_{n}  (z_0) | \prod_{i=1}^n {\cal O}^{R (p_i,q_i)}_{m_i} \rangle = \langle {\cal O}^{R (P,Q)}_{M} (\infty) {\cal O}^{R (p,q)}_{n}  (z_0)  \prod_{i=1}^n {\cal O}^{R (p_i,q_i)}_{m_i} (0) \rangle \label{mumti}
\ee
The conformal weight of the twist $n$ operator gives a Jacobian factor under this conformal map. Recall that the weights of the insertion operator are
\bee
h = \frac{1}{2} \left ( p + n - 1 \right ) \qquad
\bar{h} = \frac{1}{2} \left ( q + n - 1 \right ) 
\ee
and thus the Jacobian factor induced is 
\bee
\left ( \frac{dz}{dw} \right )^h_{| w_0} \left ( \frac{d \bar{z}}{d \bar{w}} \right ) ^{\bar{h}}_{|  \bar{w}_0}, 
\ee
which can immediately be written as 
\bee
\exp ( h w_0) \exp ( \bar{h} \bar{w}_0). \label{func1}
\ee
To rewrite this expression in terms of $t_0$, we need to use \eqref{wtmap}. Here we will be primarily interested in calculating correlation functions for which $p=q=0$ and thus the Jacobian factor gives
\bee
| t_0 |^{M(n-1)}  \left ( \prod_{i=2}^n (1 - \bar{a}_i)^{m_i} \right )^{n-1} \label{func1a}
\ee
where the $\bar{a}_i$ are defined in \eqref{bari}. 

Next we can express \eqref{mumti} in terms of a normalised correlation function:
\bee
\langle {\cal O}^{R (P,Q)}_{M} (\infty) {\cal O}^{R (p,q)}_{n}  (z_0)  \prod_{i=1}^n {\cal O}^{R (p_i,q_i)}_{m_i} (0) \rangle 
= \Lim{|z| \rightarrow \infty} \frac { \langle {\cal O}^{R (P,Q)}_{M} (z) {\cal O}^{R (p,q)}_{n}  (z_0)  \prod_{i=1}^n {\cal O}^{R (p_i,q_i)}_{m_i} (0) \rangle }{ \langle {\cal O}^{R (P,Q)}_{M} (z)  {\cal O}^{R (P,Q) \dagger}_{M} (0) \rangle }.
\ee
Here we use the notation 
\bee
{\cal O}^{R (P,Q) \dagger}_{M}
\ee
to denote the conjugate operator (with conjugate $R$ charges). 

Following \cite{Lunin:2000yv,Lunin:2001pw}, the key point is then that this normalised correlation function factorises into a bare twist part (associated with a Liouville action) and a spin field part i.e. 
\bee
\Lim{|z| \rightarrow \infty} \frac { \langle {\cal O}^{R (P,Q)}_{M} (z) {\cal O}^{R (p,q)}_{n}  (z_0)  \prod_{i=1}^n {\cal O}^{R (p_i,q_i)}_{m_i} (0) \rangle }{ \langle {\cal O}^{R (P,Q)}_{M} (z)  {\cal O}^{R (P,Q) \dagger}_{M} (0) \rangle }
= 
\Lim{|z| \rightarrow \infty} \frac { \langle \Sigma_{n+2}(z,z_0) \rangle }{ \langle \Sigma_2 (z) \rangle } \frac { \langle S_{n+2}(t,t_0) \rangle }{ \langle S_2 (t) \rangle}.
\ee
Here the bare twist part is 
\bee
 \langle \Sigma_{n+2}(z,z_0) \rangle : = \langle \sigma_{M}(z) \sigma_n (z_0) \prod_{i=1}^n \sigma_{m_i} (0) \rangle \label{t11}
\ee
with 
\bee
\langle \Sigma_2 (z) \rangle : = \langle \sigma_M (z) \sigma_M (0) \rangle. \label{t12}
\ee
The spin field correlators are calculated on the $t$ plane:
\bee
 \langle S_{n+2}(t,t_0) \rangle : = \langle S^{(P,Q)}_{M}(t(z)) S^{(p,q)}_n (t_0(z_0)) \prod_{i=1}^n S^{(p_i,q_i)}_{m_i} (a_i) \rangle \label{s11}
\ee
and
\bee
\langle S_2(t) \rangle :=  \langle S^{(P,Q)}_M (t(z)) S^{(P,Q) \dagger}_M (0) \rangle. \label{s12}
\ee
We will discuss below how the operator/state R charges (indicated in the labelling of these spin fields) relate to the spins of the spin fields. In the rest of this section we collect all the contributions required to compute the correlation function.

\subsection{Twist operator correlator}

In this section we will calculate the bare twist operator contribution, namely
\bee
{\rm Lim}_{|z| \rightarrow \infty} \frac { \langle \Sigma_{n+2}(z,z_0) \rangle }{ \langle \Sigma_2 (z) \rangle } 
\ee
where the twist operator correlators are defined in \eqref{t11} and \eqref{t12}. 

Following \cite{Lunin:2000yv,Lunin:2001pw} we will work in a path integral formulation and regularise each twist operator inserted at a finite value of $z$ by cutting out a hole of radius $\epsilon \ll 1$. The regularised twist operator $\sigma_m^{\epsilon}$  is related to the original twist operator as
\bee
\sigma_m = \frac{1}{\sqrt{\sigma^{\epsilon}_m(0) \sigma^{\epsilon}_m(1)}} \sigma_m^{\epsilon},
\ee
and thus when working with such regularised operators we need to take into account the appropriate normalisation factors. 
If a twist operator is inserted at infinity, we need to cut out a hole at infinity with radius $1/\delta \gg 1$; we denote the corresponding regularised operator as $\sigma^{\delta}_{M}$. 

Thus we need to calculate 
\bee
{\rm Lim}_{|z| \rightarrow \infty} \frac { \langle \Sigma_{n+2}(z,z_0) \rangle }{ \langle \Sigma_2 (z) \rangle } 
=  {\cal N}_{\epsilon} \frac{ \langle \sigma^{\delta}_{M}(\infty) \sigma^{\epsilon}_n (z_0) \prod_{i=1}^n \sigma^{\epsilon}_{m_i} (0) \rangle 
}{\langle \sigma^{\delta}_M (\infty) \sigma^{\epsilon}_M (0) \rangle},
\ee
where the normalisation factor is
\bee
{\cal N}_{\epsilon} = \sqrt{ \frac{ \langle \sigma^{\epsilon}_M (0) \sigma^{\epsilon}_M (1) \rangle}{
\langle \sigma^{\epsilon}_n (0) \sigma^{\epsilon}_n (1) \rangle
\prod_{i=1}^n \langle \sigma^{\epsilon}_{m_i} (0) \sigma^{\epsilon}_{m_i} (1) \rangle } }. \label{caln}
\ee
Note that normalisation terms cancel for the operator inserted at infinity. 

The two point functions of regularised twist operators at finite separation are given by \cite{Lunin:2000yv,Lunin:2001pw}
\bee
\langle \sigma^{\epsilon}_m (0) \sigma^{\epsilon}_m (y) \rangle = y^{- (m - \frac{1}{m})} \left ( m^2 e^{- \frac{(m-1)^2}{m}}\right ) Q^{1-m}
\ee
where $Q$ depends on the regularization. Factors of $Q$ cancel in the normalisation factor \eqref{caln}. Thus the overall normalisation factor is
\bee
{\cal N}_{\epsilon} = \frac{M}{n \prod_{i} m_i} \epsilon^{  - \frac{(M-1)^2}{2 M} + \frac{(n-1)^2}{2n} + \sum_i \frac{(m_i-1)^2}{2 m_i} }.
\ee

\bigskip 

The correlation functions of the regularised twist operators 
are calculated using the Liouville action associated with the conformal map from the $z$ plane to the $t$ plane. 
This conformal map changes the metric by a factor of $\exp(\phi)$ where 
\bee
\phi = \log \left | \frac{dz}{dt} \right |^2.
\ee
Under this map the Liouville contribution to the path integral reduces to boundary contributions 
\bee
S_L = \frac{c}{96 \pi} \left (i \int_{\partial \Sigma_t} \phi \partial_t \phi + {\rm c. c.} \right ) \label{liou1}
\ee
where the boundaries are the images in the $t$ plane of the circular holes cut out in the $z$ plane to regularise the operators. Here the central charge $c = 6$. 

Let us now calculate the Liouville contribution associated with the twist $m_i$ operator, for which the insertion point in the $t$ plane is $t = a_i$. In the neighbourhood of $t = a_i$ the ramification map is
\bee
z \approx (t - a_i)^{m_i} \prod_{i \neq j} (a_i-a_j)^{m_j}
\ee
and thus 
\bee
(t - a_i) \approx \left (  \frac{z}{\prod_{i \neq j} (a_i -a_j)^{m_j}} \right )^{\frac{1}{m_i}}.
\ee
Thus the Liouville field in the vicinity of $t = a_i$ is given by 
\bee
\phi \approx 2 \log \left ( m_i | t - a_i|^{m_i-1} \prod_{i \neq j} | a_i - a_j |^{m_j} \right )
\ee
with 
\bee
\partial_t \phi \approx \frac{m_i - 1}{(t -a_i)}.
\ee
The contribution to the Liouville action from this point is given by \eqref{liou1}, with the integral evaluated using 
\bee
z \approx \epsilon e^{i \theta}, \qquad
(t  - a_i) \approx  \left ( \frac{\epsilon}{\prod_{i \neq j} (a_i - a_j)^{m_j}} \right )^{\frac{1}{m_i}} e^{i \theta'}, \qquad
\theta' = \frac{\theta}{m_i},
\ee
where the range of $\theta'$ is $2 \pi$. Thus the contribution from $t = a_i$ is
\bee
S_L ^{a_i} = - \frac{1}{2} (m_i  -1)  \log \left (m_i \epsilon^{\frac{m_i-1}{m_i}} \prod_{i \neq j} | a_i - a_j |^{\frac{m_j}{m_i}}  \right ).
\ee
We now consider the contribution from the point associated with the twist $n$ operator. In the neighbourhood of the insertion point
\bee
z - z_0 \approx b_n ( t - t_0)^n \label{bn-def}
\ee
where the coefficient $b_n$ will be calculated in the next subsection. Following the same logic as above, we can immediately 
write down the associated contribution to the Liouville action
\bee
S_L^{t_0} = - \frac{1}{2 n} (n- 1) \log \left ( n^n | b_n | \epsilon^{n-1} \right ).
\ee
For the insertion at infinity
\bee
z \approx t^{M}
\ee
and the Liouville action contribution is 
\bee
S_L^{\infty} = \frac{1}{2} (M -1) \log \left ( M \delta^{-\frac{M-1}{M}} \right ). \label{l-inft}
\ee
Note that the opposite sign relative to the previous contributions follows from the direction of the boundary normal. 

Collecting together all of these contributions we obtain
\bea
S^{(4)}_L &=& - \sum_i \frac{1}{2} (m_i  -1)  \log \left (m_i \epsilon^{\frac{m_i-1}{m_i}} \prod_{i \neq j} | a_i - a_j |^{\frac{m_j}{m_i}}  \right ) \nonumber \\
&& - \frac{1}{2 n} (n- 1) \log \left ( n^n | b_n | \epsilon^{n-1} \right ) + \frac{1}{2} (M -1) \log \left ( M \delta^{-\frac{M-1}{M}} \right ).
\eea
The regularised four point function is now calculated as 
\bee
\langle \sigma^{\delta}_{M}(\infty) \sigma^{\epsilon}_n (z_0) \prod_{i=1}^n \sigma^{\epsilon}_{m_i} (0) \rangle = e^{S^{(4)}_L}.
\ee
The calculation of the regularised two point function 
\bee
\langle \sigma^{\delta}_{M}(\infty)  \sigma^{\delta}_{M}(0) \rangle = e^{S^{(2)}_L}
\ee
is very similar. The Liouville contribution from the insertion at infinity is given by \eqref{l-inft} and the contribution at zero is
\bee
S_L^{\infty} = - \frac{1}{2} (M -1) \log \left ( M \epsilon^{-\frac{M-1}{M}} \right ).
\ee
Thus the total Liouville action contribution to the two point function is 
\bee
S^{(2)}_L =  \frac{1}{2} (M -1) \log \left ( M \delta^{-\frac{M-1}{M}} \right ) - \frac{1}{2} (M -1) \log \left ( M \epsilon^{-\frac{M-1}{M}} \right ).
\ee
Collecting all of the holomorphic and anti-holomorphic contributions together we ultimately obtain
\bee
M^{\frac{1}{2} (M+1)} 
n^{-\frac{1}{2} (n+1)} |b_n|^{- \frac{(n-1)}{2n}} \prod_{i=1}^n m_i^{-\frac{1}{2} (m_i + 1)} \prod_{i \neq j} |a_i - a_j |^{-\frac{m_j (m_i - 1)}{2 m_i}}, \label{func2}
\ee
where implicitly we set $a_1 = 0$. Note that all contributions depending on the regulators $\epsilon$ and $\delta$ cancel, as required. 

\subsection{Spin field correlator}

In this section we calculate 
\bee
{\rm Lim}_{|z| \rightarrow \infty}  \frac { \langle S_{n+2}(t(z),t_0(z_0)) \rangle }{ \langle S_2 (t(z)) \rangle},
\ee
where the correlators are defined in \eqref{s11} and \eqref{s12}. 

The relationship between the R charge assignments of the original operator/states and the spin field labels is as follows. The operator creating a component string of twist $m$ is mapped to
\bee
{\cal O}_m^{R(p,q)} \rightarrow S_m^{(p,q)} \sigma_{m}
\ee
where $\sigma_m$ is the bare twist $m$ operator and $S_m^{(p,q)}$ has $SU(2)_L$ and $SU(2)_R$ charges
\bee
\frac{1}{2} (p-1) \qquad
\frac{1}{2} (q-1). \label{slabel1}
\ee
For the twist $n$ operator, the mapping is 
\bee
{\cal O}_n^{R(p,q)} \rightarrow S_n^{(p,q)} \sigma_{n}
\ee 
with the $SU(2)_L$ and $SU(2)_R$ charges of $S_n^{(p,q)}$ being
\bee
\frac{1}{2} (p + n -1) \qquad
\frac{1}{2} (q + n -1). \label{slabel2}
\ee
Note that the correlation function calculations in \cite{Lunin:2001pw} are applicable to universal operators common to both the $T^4$ and K3 CFTs, i.e. operators associated with the $(0,0)$ and $(2,2)$ cohomology.

As a warm up  we will consider an example of twist three operator joining three components; the case of a twist two operator joining two components can be found in \cite{Carson:2014yxa}. We consider R charge assignments such that we need to calculate
\bee
\langle S_{5}(t,t_0) \rangle : = \langle S^{(\frac{1}{2},\frac{1}{2})}_{M}(t(z)) S^{(1,1)}_3 (t_0(z_0)) \prod_{i=1}^3 S^{(- \frac{1}{2},- \frac{1}{2})}_{m_i} (a_i) \rangle \label{s13}
\ee
and
\bee
\langle S_2(t) \rangle :=  \langle S^{(\frac{1}{2},\frac{1}{2})}_M (t(z)) S^{(-\frac{1}{2},-\frac{1}{2}) }_M (0) \rangle. \label{s14}
\ee
Thus, the original one point function involves only operators associated with the $(0,0)$ cohomology. 

We begin by collecting the normalisation factors for the spin fields. For a spin field associated with a twist $m$ operator, the ramification map by construction takes the form 
\bee
(z - z_m) = b_m (t - t_m)^m 
\ee
near the insertion point $z_m$ (mapped to $t_m$). The corresponding (holomorphic) normalisation factor for the spin field insertion is then
\bee
b_m^{-\frac{j_3^2}{m}},
\ee
where $j_3$ is the $SU(2)_L$ charge of the spin field. Here and throughout this section we explain in detail the holomorphic contributions; we then combine the holomorphic and anti-holomorphic factors to obtain the full result.

For the component strings this results in normalisation factors
\bea
b_{m_1} &=& (-a_2)^{m_2} (-a_3)^{m_3} \qquad
b_{m_1}^{-\frac{1}{4 m_1}} = (-a_2)^{- \frac{m_2}{4 m_1}} (-a_3)^{-\frac{m_3}{4m_1}} \nonumber \\
b_{m_2} &=& a_2^{m_1} (a_2 - a_3)^{m_3} \qquad
b_{m_2}^{-\frac{1}{4 m_2}} = a_2^{- \frac{m_1}{4 m_2}} (a_2 - a_3)^{-\frac{m_3}{4m_2}} \nonumber \\
b_{m_3} &=& a_3^{m_1} (a_3 - a_2)^{m_2} \qquad
b_{m_3}^{-\frac{1}{4m_3}} = a_3^{-\frac{m_1}{4m_3}} (a_3 - a_2)^{-\frac{m_2}{4 m_3}}.
\eea
Taking the product of these factors we obtain
\bee
t_0^{\frac{3}{4} - \frac{M}{4} \left ( \frac{1}{m_1} + \frac{1}{m_2} + \frac{1}{m_3} \right ) } (- \bar{a}_2)^{- \frac{m_2}{4 m_1}} (-\bar{a}_3)^{-\frac{m_3}{4m_1}}
\bar{a}_2^{- \frac{m_1}{4 m_2}} (\bar{a}_2 - \bar{a}_3)^{-\frac{m_3}{4m_2}}
\bar{a}_3^{-\frac{m_1}{4m_3}} (\bar{a}_3 - \bar{a}_2)^{-\frac{m_2}{4 m_3}}.  \label{comb1}
\ee
For the twist three operator joining these component strings we find that 
\bee
b_3 = \frac{M}{3} t_0^{M-3} (1 - \bar{a}_2)^{m_2 -1} (1 - \bar{a}_3)^{m_3 - 1}
\ee
and thus this spin operator normalisation is 
\bee
t_0^{1 - \frac{M}{3}} \left ( \frac{M}{3} (1 - \bar{a}_2)^{m_2 -1} (1 - \bar{a}_3)^{m_3 - 1} \right )^{\frac{1}{3}}. \label{comb2}
\ee
The normalisation factors from the twist $M$ operator are trivial in both the five point function and the two point function since $b_M = 1$. 
The combination of \eqref{comb1} and \eqref{comb2} results in a term proportional to:
\bee
t_0^{\frac{7}{4} - M \left ( \frac{1}{3} + \frac{1}{4} \sum_{i}\frac{1}{m_i} \right )}.
\ee
The complete normalisation factor is obtained by combining both the holomorphic and anti-holomorphic parts, leading to a term proportional to:
\bee
|t_0|^{\frac{7}{2} - M \left ( \frac{2}{3} + \frac{1}{2} \sum_{i}\frac{1}{m_i} \right )}.
\ee

\bigskip 

Having studied the case of three strings being joined, it is straightforward to generalise to the joining of $n$ strings. The normalisation factors give a contribution of 
\bee
b_{n}^{-\frac{(n-1)^2}{4n}} \prod_{i=1}^n b_{m_i}^{-\frac{1}{4 m_i}} \label{opnorm}
\ee
where 
\bee
b_{m_i} = \prod_{j \neq i} (a_i - a_j)^{m_j}.
\ee
In \eqref{opnorm} we have used the fact that the R charge of the twist $n$ operator is $(n-1)/2$. 
Thus the complete normalisation factor from holomorphic and anti-holomorphic parts gives
\bee
| b_n |^{-\frac{(n-1)^2}{2n}} \prod_{i=1}^n  \prod_{j \neq i} |a_i - a_j |^{-\frac{m_j}{2 m_i}}. \label{func3}
\ee
We can calculate $b_n$ explicitly as follows. From the definition of $b_n$ in \eqref{bn-def} it is clear that close to $t_0$
\bee
\left ( \frac{dz}{dt} \right ) \approx n b_n (t - t_0)^{n-1}.  \label{bn1}
\ee
Differentiating the ramification map directly gives 
\bee
\left ( \frac{dz}{dt} \right ) \approx \prod_i (t_0 - a_i)^{m_i -1} \left ( M t^{n-1} + \cdots \right ). \label{bn2}
\ee
Comparing \eqref{bn1} and \eqref{bn2} gives
\bee
b_n = \frac{M}{n} t_0^{M-n} \prod_i (1 - \bar{a}_i)^{m_i  -1}, \label{bn-final}
\ee
where we use the dimensionless quantities $\bar{a}_i$ to make the $t_0$ dependence of $b_n$ manifest. 

\bigskip

Let us now move to the spin field correlators. Each of the spin fields factorises into holomorphic and antiholomorphic fields i.e. we can write
\bee
S_{m}^{(p,q)} = S^{(j_3)}(t) \bar{S}^{(\bar{j}_3)} (\bar{t})
\ee
where the $SU(2)_{L/R}$ charges are given in \eqref{slabel1} and \eqref{slabel2}. 

As above, let us consider first the case in which three component strings are joined, before moving on to the general case. 
For operators associated with the $(0,0)$ cohomology in the holomorphic sector we therefore need to calculate
\bee
\frac{ \langle S^{(\frac{1}{2})} (t) S^{(1)} (t_0) S^{ (- \frac{1}{2})} (0) S^{ (- \frac{1}{2} )} (a_2) S^{ (-\frac{1}{2} )} (a_3) \rangle}
{\langle S^{ ( \frac{1}{2} )} (t) S^{ (-\frac{1}{2})} (0) \rangle}
\ee
in the limit that $t \rightarrow \infty$. This correlation function can be computed following the methods of \cite{Lunin:2001pw}.

Using bosonisation we can write the spin fields as
\bee
S^{ \frac{k}{2}} (t)  = \exp \left ( \frac{i k}{2} (\phi_1 (t) - \phi_2(t)) \right ).
\ee
The OPE for these fields is 
\bee
\exp \left ( i k_1 \phi (t) \right ) \exp \left ( i k_2 \phi(t') \right ) \sim \exp \left ( i k_1 \phi(t) + i k_2 \phi(t') \right ) (t - t')^ {k_1 k_2}.
\ee
Using this OPE to compete the two point function and five point function (with appropriate ordering) we then obtain 
\bea
{\langle S^{ ( \frac{1}{2} )} (t) S^{ (-\frac{1}{2})} (0) \rangle} &=& \frac{1}{t^{\frac{1}{2}}} \\
\langle S^{(\frac{1}{2})} (t) S^{(1)} (t_0) S^{ (- \frac{1}{2})} (0) S^{ (- \frac{1}{2} )} (a_2) S^{ (-\frac{1}{2} )} (a_3) \rangle
&=&  \frac{ (t-t_0) (-a_2)^{\frac{1}{2}} (-a_3)^{\frac{1}{2}} (a_2 - a_3)^{\frac{1}{2}}}{  t^{\frac{1}{2}}(t - a_2)^{\frac{1}{2}} (t- a_3)^{\frac{1}{2}} 
t_0 (t_0 - a_2)(t_0-a_3)}. \nonumber
\eea
Thus as $t \rightarrow \infty$
\bee
\frac{ \langle S^{(\frac{1}{2})} (t) S^{(1)} (t_0) S^{ (- \frac{1}{2})} (0) S^{ (- \frac{1}{2} )} (a_2) S^{ (-\frac{1}{2} )} (a_3) \rangle}
{\langle S^{ ( \frac{1}{2} )} (t) S^{ (-\frac{1}{2})} (0) \rangle} \rightarrow \frac{ (-\bar{a}_2)^{\frac{1}{2}} (-\bar{a}_3)^{\frac{1}{2}} (\bar{a}_2 - \bar{a}_3)^{\frac{1}{2}}}{ 
t_0^{\frac{3}{2}} (1 - \bar{a}_2)(1 - \bar{a}_3)}.
\ee
Combining holomorphic and anti-holomorphic contributions we obtain
\bee
{\rm Lim}_{|z| \rightarrow \infty}  \frac { \langle S_{5}(t(z),t_0(z_0)) \rangle }{ \langle S_2 (t(z)) \rangle}
 \frac { \langle \bar{S}_{5}(\bar{t}(\bar{z}),\bar{t}_0(\bar{z}_0)) \rangle }{ \langle \bar{S}_2 (\bar{t}(\bar{z})) \rangle} =
\frac{ | \bar{a}_2| | \bar{a}_3|  | \bar{a}_2 - \bar{a}_3|}{ 
|t_0|^3 | 1 - \bar{a}_2|^2 |1 - \bar{a}_3|^2}
\ee
as the final result for the spin field correlator contribution. 

\bigskip

The generalisation to twist $n$ operators joining $n$ component strings is now immediate.  Following \eqref{s13} we choose R charge assignments such that 
\bee
\langle S_{n+2}(t,t_0) \rangle : = \langle S^{(\frac{1}{2},\frac{1}{2})}_{M}(t(z)) S^{(\frac{1}{2}(n-1),\frac{1}{2}(n-1))}_n (t_0(z_0)) \prod_{i=1}^n S^{(- \frac{1}{2},- \frac{1}{2})}_{m_i} (a_i) \rangle \label{s1n}
\ee
(with $a_1 = 0$). Then
\bee
\langle S_{n+2}(t,t_0) \rangle = \frac{ (t-t_0)^{\frac{1}{2}(n-1)} \prod_{i=1}^{n} \prod_{i < j}  (a_i -a_j)^{\frac{1}{2}}}
{  \prod_{i=1}^{n} (t - a_i)^{\frac{1}{2}} (t_0 - a_i)^{\frac{1}{2} (n-1)}} 
\ee
and thus the normalised correlator is
\bee
\frac{\langle S_{n+2}(t,t_0) \rangle }{\langle S^{ ( \frac{1}{2} )} (t) S^{ (-\frac{1}{2})} (0) \rangle} = 
\frac{ \prod_{i=1}^n \prod_{i < j}  (a_i -a_j)^{\frac{1}{2}}}
{\prod_{i=1}^n (t_0 - a_i)^{\frac{1}{2} (n-1)}}.
\ee
Combining holomorphic and anti-holomorphic contributions we obtain
\bee
{\rm Lim}_{|z| \rightarrow \infty}  \frac { \langle S_{n+2}(t(z),t_0(z_0)) \rangle }{ \langle S_2 (t(z)) \rangle}
 \frac { \langle \bar{S}_{n+2}(\bar{t}(\bar{z}),\bar{t}_0(\bar{z}_0)) \rangle }{ \langle \bar{S}_2 (\bar{t}(\bar{z})) \rangle} =\frac{ \prod_{i=1}^{n}  \prod_{i < j}  | \bar{a}_i - \bar{a}_j |}
{| t_0|^{\frac{1}{2}  n (n-1)} \prod_{i=2}^n | 1 - \bar{a}_i |^{ (n-1)} } \label{func4}
\ee
for the spin field correlator associated with the given R charge assignments. 

\subsection{Final answer for one point function}

The final answer for the one point function is obtained by combining \eqref{func1a}, \eqref{func2}, \eqref{func3} and \eqref{func4}. 
First note that combining \eqref{func2} and \eqref{func3} gives
\bee
M^{\frac{1}{2} (M+1)} 
n^{-\frac{1}{2} (n+1)} |b_n|^{- \frac{1}{2}(n-1) } \prod_{i=1}^n m_i^{-\frac{1}{2} (m_i + 1)} \prod_{i \neq j} |a_i - a_j |^{-\frac{m_j }{2}} 
\ee
which can be rewritten as 
\bee
M^{\frac{1}{2} (M+1)} 
n^{-\frac{1}{2} (n+1)} |b_n|^{- \frac{1}{2} (n-1)} |t_0|^{- \frac{1}{2} M(n-1)} \prod_{i=1}^n m_i^{-\frac{1}{2} (m_i + 1)} \prod_{i \neq j} | \bar{a}_i - \bar{a}_j |^{-\frac{m_j }{2}}. 
\ee
Substituting the expression for $b_n$ from \eqref{bn-final}, the $t_0$ dependence is
\bee
| t_0 |^{\frac{1}{2} n (n - 1) - M (n-1)}.
\ee
Since the $t_0$ dependence of \eqref{func1a} is $| t_0|^{M(n-1)}$ and the $t_0$ dependence of \eqref{func4} is $ | t_0|^{- \frac{n(n-1)}{2}}$, all factors of $t_0$ cancel from the final result, as expected. 

Combining the remaining terms in \eqref{func1a}, \eqref{func2}, \eqref{func3} and \eqref{func4} gives
\bea 
&& \langle {\cal O}^{R (0,0))}_{M}| {\cal O}^{(0,0)}_{n}  | \prod_{i=1}^n {\cal O}^{R (0,0)}_{m_i} \rangle = \label{final0}  \\
&& \qquad \qquad  \frac{M^{\frac{1}{2} (M+ 2 - n)}}{n} \prod_i | 1- \bar{a}_i |^{\frac{1}{2} (m_i - 1)(n-1)} \prod_j m_j^{-\frac{1}{2} (m_j + 1)} \prod_{j \neq k} | \bar{a_j} - \bar{a}_k |^{\frac{1}{2} (1-m_k)}, \nonumber
\eea
where we recall that $\bar{a}_i$ is defined in \eqref{bari}. 

\subsection{Special cases: \texorpdfstring{$n=2$}{n=2} and \texorpdfstring{$n=3$}{n=3}}

In this section we consider the limit of this correlation function in special cases. We begin with the case of $n=2$, which was already studied in \cite{Carson:2014yxa}. In this case the correlation function reduces to the simple expression
\bee
\langle {\cal O}^{R (0,0))}_{M}| {\cal O}^{(0,0)}_{2}  (w_0)  | {\cal O}^{R (0,0)}_{m_1} {\cal O}^{R (0,0)}_{m_2} \rangle = \frac{M}{2 m_1 m_2},
\ee
in agreement with \cite{Carson:2014yxa}. 

Now let us turn to the case of $n=3$. The correlator \eqref{final0} can in this case be written as 
\bea
&& \frac{1}{3} M^{\frac{1}{2} (M-1)} (1 - \bar{a}_2)^{m_2 - 1} (1 - \bar{a}_3)^{m_3 - 1} m_1^{-\frac{1}{2} (m_1 +1)} m_2^{-\frac{1}{2} (m_2 + 1)}
m_3^{-\frac{1}{2} (m_3 + 1} \\
&&| \bar{a}_2 |^{1 - \frac{1}{2} (m_1 + m_2)} | \bar{a}_3 |^{1 - \frac{1}{2} (m_1 + m_3)} | \bar{a}_2 - \bar{a}_3 |^{1 - \frac{1}{2} (m_2 + m_3)}. \nonumber
\eea
Note that this expression appears asymmetric between the twist $m_i$ operators only because we have set $\bar{a}_1 = 0$; the expression could trivially be symmetrised by reinstating the $\bar{a}_1$ terms. 

This expression looks extremely complicated but in fact it simplifies to give a very concise result. Using 
\bea
| \bar{a}_2 | &=& \frac{M^{\frac{1}{2}}}{m_1^{\frac{1}{2}} (m_1 + m_3)} \sqrt{m_1 M + m_2 m_3} \nonumber \\
| \bar{a}_3 | &=& \frac{M^{\frac{1}{2}}}{m_1^{\frac{1}{2}} (m_1 +m_2)} \sqrt{m_1 M + m_2 m_3}
\eea
and 
\bea
| 1 - \bar{a}_2 | &=& \frac{m_2^{\frac{1}{2}}}{ m_1^{\frac{1}{2}} (m_1 + m_3)} \sqrt{m_3 M + m_1 m_2} \nonumber \\ 
| 1 - \bar{a}_3 | &=& \frac{m_3^{\frac{1}{2}}}{m_1^{\frac{1}{2}} (m_1 + m_2)} \sqrt{m_2 M + m_1 m_3}
\eea
and
\bee
| \bar{a}_2 - \bar{a}_3 | = \frac{M^{\frac{1}{2}}}{m_1^{\frac{1}{2}} (m_1 + m_2) (m_1 + m_3)} \sqrt{m_3 M + m_1 m_2} \sqrt{m_2 M + m_1 m_3}
\ee
together with relations such as
\bee
(m_1 M + m_2 m_3) = (m_1 + m_2) (m_1 + m_3)
\ee 
we can show that the correlator simplifies to 
\bee
\langle {\cal O}^{R (0,0))}_{M}| {\cal O}^{(0,0)}_{3}  (w_0)  | {\cal O}^{R (0,0)}_{m_1} {\cal O}^{R (0,0)}_{m_2} {\cal O}^{R (0,0)}_{m_3} \rangle = \frac{M}{3 m_1 m_2 m_3}.
\ee
This expression is manifestly symmetric over the $m_i$; furthermore, all dependence on factors of the type $(m_i + m_j)$ cancels out. 

Given the special cases considered in this section, it would be natural to conjecture that the result for the general case is
\bee
\langle {\cal O}^{R (0,0))}_{M}| {\cal O}^{(0,0)}_{n}  (w_0)  | \prod_{i} {\cal O}^{R (0,0)}_{m_i} \rangle = \frac{M}{n \prod_{i} m_i}, \label{conj}
\ee
but this is not  supported by the results below.

\subsection{Case of equal \texorpdfstring{$m_i$}{mi}}

In this section we consider the case in which the $n$ strings are of equal length i.e. $m_i = M/n$. In this case the general expression \eqref{final0} simplifies considerably to
\bee
M^{1-n} n^{\frac{1}{2} M + \frac{1}{2} n - 1} \prod_{i=2}^{n} |1 - \bar{a}_i |^{\frac{1}{2}(\frac{M}{n} - 1)(n-1)} \prod_{j=1}^n 
\prod_{j \neq k} | \bar{a}_j - \bar{a}_k |^{\frac{1}{2} \left(1 - \frac{M}{n} \right )}. \label{final01}
\ee
The zeroes of the ramification map are located at:
\bee
\bar{a}_i = \frac{n}{(n-1)} \left ( 1 + \bar{a} \exp (i \phi_i) \right ), \qquad i \ge 2 
\ee
where 
\bee
\bar{a}^{n-1} = n \left ( 1 - \frac{1}{n} \right )^{n-1} - 1
\ee
and the phases are given as before by \eqref{phase-sol}. Using the properties of the phases we can then immediately show that 
\bee
\prod_{j = 2}^{n} |  \bar{a}_j |^{\frac{1}{2} (1 - \frac{M}{n})} = n^{ \frac{1}{2} \left (1 - \frac{M}{n} \right )} \label{prev}
\ee
and hence we can write \eqref{final01} as 
\bee
M^{1-n} n^{\frac{1}{2} M + \frac{1}{2} n - \frac{M}{n}} \prod_{i=2}^{n} |1 - \bar{a}_i |^{\frac{1}{2}(\frac{M}{n} - 1)(n-1)} \prod_{j=2}^n 
\prod_{j \neq k} | \bar{a}_j - \bar{a}_k |^{\frac{1}{2} \left(1 - \frac{M}{n} \right )}, \label{final1}
\ee
i.e. we can immediately evaluate the products involving $a_1 = 0$. (Note that this evaluation gives \eqref{prev} squared.)

To evaluate \eqref{final1} we make use of 
\bee
(1 - \bar{a}_i ) = - \frac{1}{(n-1)} \left (1 + n \bar{a} \exp (i \phi_i) \right )
\ee
and
\bee
\bar{a}_i - \bar{a}_j = \frac{n \bar{a}}{(n-1)} \left ( \exp(i \phi_i) - \exp( i\phi_j) \right ).
\ee
Note that the latter expression has a geometric interpretation: the $(n-1)$ ramification zeroes $\{ a_i \}$ are located at the vertices of a regular $(n-1)$ polygon in the complex plane. The expression
\bee
v_{ij} := \left ( \exp(i \phi_i) - \exp( i \phi_j) \right )
\ee
can thus be interpreted vectorially in terms of vectors connecting the vertices of such a regular $(n-1)$ polygon, in which the vertices are unit distance from the origin of the complex plane.
We represent this in figure~\ref{fig:dodecagon}.

\begin{figure}[tbp]
\centering
\includegraphics[width=0.5\linewidth]{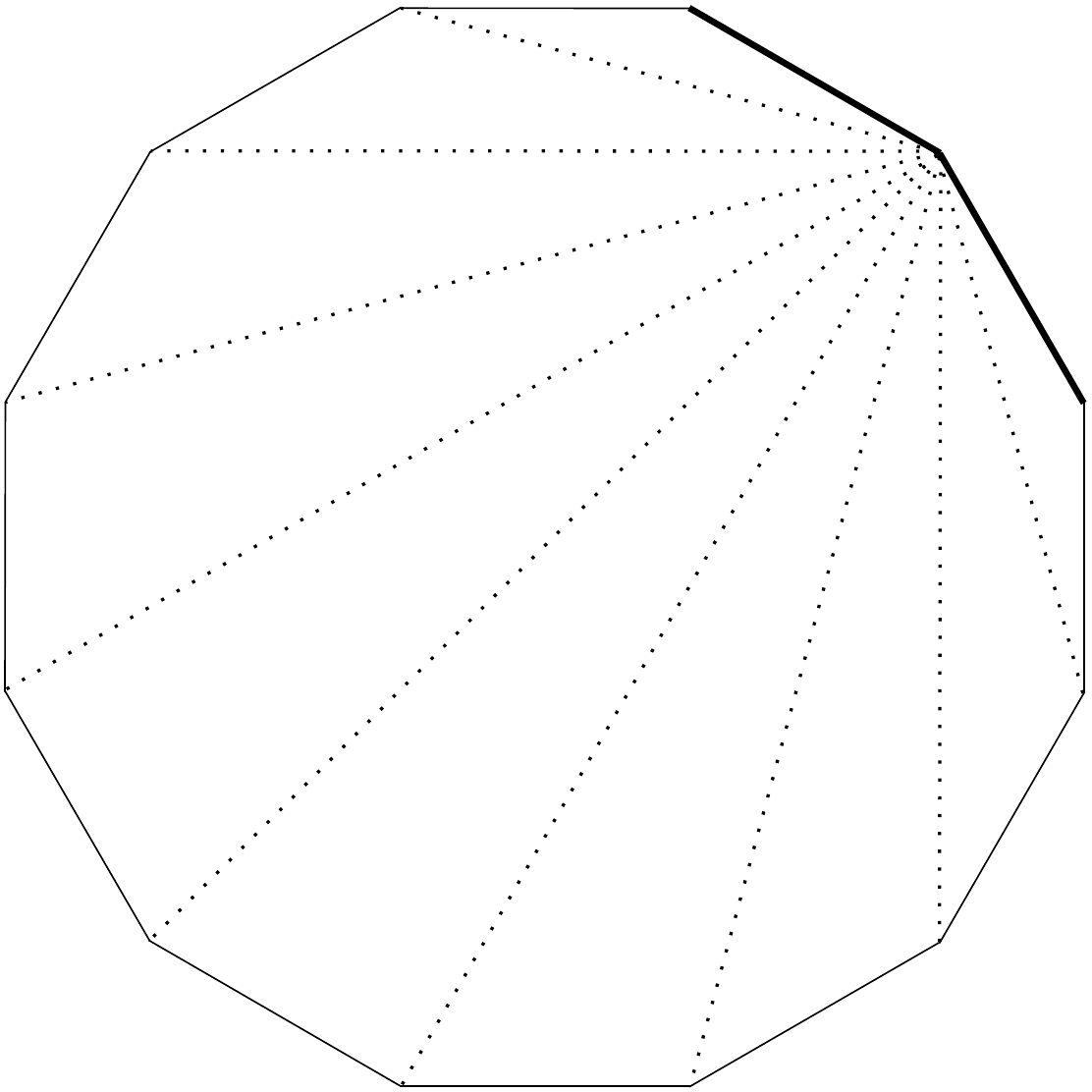}
\caption{Polygon representing the location of the ramification zeroes. We represent with dotted lines all the diagonals of one vertex, and with thicker lines its two adjacent sides.}
\label{fig:dodecagon}
\end{figure}

The first product of \eqref{final1} can be written as 
\bee
\prod_{i=2}^n \left (1 - \bar{a}_i \right ) = \frac{(-1)^{n-1}}{(n-1)^{(n-1)}} \left ( 1 + n \bar{a} \sum_{i=2}^{n} \exp(i \phi_i) + \cdots + 
(n \bar{a})^{n-1} \prod_{i=2}^n \exp (i \phi_i) \right ).
\ee
Using the properties of the phases \eqref{phase-sol} we can then show that this reduces to 
\bee
\prod_{i=2}^n \left (1 - \bar{a}_i \right ) = \frac{(-1)^{n-1}}{(n-1)^{(n-1)}} \left (1 + n^{n-1} \bar{a}^{n-1} \right )
\ee
and thus 
\bee
\prod_{i=2}^n | 1 - \bar{a}_i | = \frac{1}{(n-1)^{(n-1)}} \left (1 + n (n-1)^{(n-1)}  - n^{(n-1)} \right ). 
\ee
Hence we can evaluate the following contribution to the one point function:
\bee
\prod_{i=2}^{n} |1 - \bar{a}_i |^{\frac{1}{2}(\frac{M}{n} - 1)(n-1)} = (n-1)^{- \frac{1}{2}(\frac{M}{n} - 1)(n-1)^2} 
\left (1 + n (n-1)^{(n-1)}  - n^{(n-1)} \right )^{\frac{1}{2}(\frac{M}{n} - 1)(n-1)}.
\ee
It is more subtle to find a closed form expression for 
\bee
\prod_{j=2}^n  \prod_{j \neq k} | \bar{a}_j - \bar{a}_k | = \left ( \frac{ n \bar{a}}{(n-1)} \right )^{(n-1)(n-2)} \prod_{j=2}^n  \prod_{j \neq k} |  \exp(i \phi_j) - \exp( i \phi_k)  |
\ee
as this requires 
\bee
\prod_{j=2}^n  \prod_{j \neq k} |  \exp(i \phi_j) - \exp( i \phi_k)  |, \label{poly}
\ee
i.e. the square of the product of the side lengths and all the diagonals of the regular polygon. 
The polygon side length is given by 
\bee
2 \sin \left ( \frac{\pi}{(n-1)} \right )
\ee
while the lengths of the diagonals are given by
\bee
2 \sin \left ( \frac{ j \pi}{(n-1)} \right ) \qquad 2 \le j \le (n-2). 
\ee
Now consider a specific vertex of the regular polygon. For this vertex the total product of side lengths and diagonals is, using the two previous results
\bee
2^{(n-2)} \prod_{j=1}^{(n-2)}  \sin \left ( \frac{ j \pi}{(n-1)} \right ) = (n-1), 
\ee
where in evaluating this expression we use standard trigonometry identities. 

The polygon has in total $(n-1)$ vertices and thus we obtain
\bee
\prod_{j=2}^n  \prod_{j \neq k} |  \exp(i \phi_j) - \exp( i \phi_k)  | = (n-1)^{(n-1)}. \label{poly2}
\ee
Collecting together all the contributions we obtain
\bea
&& \langle {\cal O}^{R (0,0))}_{M}| {\cal O}^{(0,0)}_{n}   | \left ( {\cal O}^{R (0,0)}_{M/n} \right )^n \rangle = \label{equalm}  \nonumber \\
&& \qquad \qquad M^{1-n} n^{\frac{M}{2} + \frac{n}{2} - \frac{M}{n}} \left (  (n-1)^2 (n \bar{a})^{(n-2)} \Lambda^{-1} \right )^{\frac{1}{2} (n-1) \left (1 - \frac{M}{n} \right )}
\eea
where we introduced the notation 
\bee
\Lambda = \left ( 1 + n (n-1)^{(n-1)} - n^{n-1} \right ). 
\ee
Note that this does not take the simple form conjectured above \eqref{conj}, except for $n=2$ and $n=3$. 

\bigskip

It is useful to work out the expressions explicitly for low values of $n$. For {\bf $n=3$}, 
\bee
|1 - \bar{a}_2 | = | 1 - \bar{a}_3 | = 1
\ee
and 
\bee
| \bar{a}_2 - \bar{a}_3 | = \sqrt{3}. 
\ee
Combining the factors, the correlation function thus reduces to 
\bee
\langle {\cal O}^{R (0,0))}_{M}| {\cal O}^{(0,0)}_{3}  (w_0)  | {\cal O}^{R (0,0)}_{\frac{M}{3}} {\cal O}^{R (0,0)}_{\frac{M}{3}} {\cal O}^{R (0,0)}_{\frac{M}{3}} \rangle = \left ( \frac{M}{3} \right )^{-2},
\ee
in agreement with the direct limit of the expression \eqref{final0}. 

Now let us consider {\bf $n=4$}. The regular polygon used to calculate \eqref{poly} is an equilateral triangle with circumradii equal to one. Elementary geometry gives the length of the triangle side as $\sqrt{3}$ and thus \eqref{poly} reduces to $3^3$, in agreement with \eqref{poly2}. In this case
\bee
\bar{a}^3 = \frac{11}{16} 
\ee
and thus
\bee
\langle {\cal O}^{R (0,0))}_{M}| {\cal O}^{(0,0)}_{4}  (w_0)  | \left ( {\cal O}^{R (0,0)}_{\frac{M}{4}} \right )^4 \rangle = \left ( \frac{4}{M} \right )^{3} 5^{\frac{3M}{8} - \frac{3}{2}} 11^{1 - \frac{M}{4}}.
\ee
The conjecture \eqref{conj} would instead give
\bee
\langle {\cal O}^{R (0,0))}_{M}| {\cal O}^{(0,0)}_{4}  (w_0)  | \left ( {\cal O}^{R (0,0)}_{\frac{M}{4}} \right )^4 \rangle = \left ( \frac{4}{M} \right )^{3}
\ee
and therefore this simple form for the one point function cannot be correct for $n > 3$. 


\bigskip

We can also take the large $n$ limit of \eqref{equalm}. In the limit of $n \gg 1$
\bee
\Lambda \approx n^n
\ee
while 
\bee
\bar{a}^{n-1} \approx \frac{n}{e}. 
\ee
The latter follows from the limit 
\bee
\left ( 1 - \frac{1}{n} \right )^{n-1} \rightarrow \frac{1}{e}
\ee
for large $n$. Then
\bee
\langle {\cal O}^{R (0,0))}_{M}| {\cal O}^{(0,0)}_{n}  (w_0)  | \left ( {\cal O}^{R (0,0)}_{\frac{M}{n}} \right )^n \rangle \approx \left ( \frac{n}{M} \right )^{n} n^{-\frac{M}{2n}} e^{\frac{1}{2} (M-n)}.
\ee
In this expression we do not make any assumptions about the twist of the component strings, i.e. the ratio $M/n$, beyond the fact that it is a positive integer. 

\section{Conclusions and outlook} \label{section:conc}

The main result of this paper is a general expression for the amplitude for joining $n$ strings using a twist operator \eqref{final0}. As discussed in section \ref{section:twist}, this amplitude can be used to compute one point functions for supergravity operators in 2-charge and 3-charge black hole microstates. While the black hole microstate programme was the main motivation for the current work, correlation functions in the orbifold SCFT are interesting in a number of other contexts. 

In the early days of AdS/CFT, the spectrum and cubic couplings for six-dimensional ${\cal N} = 4b$ supergravity were calculated \cite{Deger:1998nm,Arutyunov:2000by}; these allowed the spectrum of chiral primary operators and three point functions of chiral primaries to be calculated. The correlation functions discussed here could be matched with higher point functions from the supergravity side, although this would require higher point supergravity interactions to be computed. 

The holographic duality for AdS$_3 \times$ S$^3 \times$ S$^3 \times$ S$^1$ was for a long time mysterious, with conjectures for the corresponding SCFT with large ${\cal N} =4$ supersymmetry discussed in  \cite{deBoer:1999gea,Gukov:2004ym}. There has recently been considerable progress on this subject, see \cite{Gaberdiel:2016xwo,Eberhardt:2017fsi,Eberhardt:2017pty,Gaberdiel:2017oqg}, with the holographic duals being conjectured to be symmetric orbifolds of minimal models. The supergravity spectrum was computed in detail in \cite{Eberhardt:2017fsi}, to match with the dual SCFT. Integrability was also used to study the spectrum in \cite{Baggio:2017kza}. The techniques of this paper would be relevant to computing correlation functions in the orbifold CFT to match the holographic correlation functions.

\section*{Acknowledgements}

This work is funded by the STFC grant ST/P000711/1. This project has received funding and support from the European Union's Horizon 2020 research and innovation programme under the Marie Sklodowska-Curie grant agreement No 690575. MMT would like to thank the Kavli Institute for the Physics and Mathematics of the Universe for hospitality during the completion of this work.

\bibliographystyle{JHEP}
\bibliography{refs}

\end{document}